\documentclass[12pt]{article}

\usepackage{scicite,graphicx}

\usepackage{times}

\newenvironment{sciabstract}{%
\begin{quote} \bf}
{\end{quote}}

\newcounter{lastnote}
\newenvironment{scilastnote}{%
\setcounter{lastnote}{\value{enumiv}}%
\addtocounter{lastnote}{+1}%
\begin{list}%
{\arabic{lastnote}.}
{\setlength{\leftmargin}{.22in}}
{\setlength{\labelsep}{.5em}}}
{\end{list}}

\title{Nonlocal edge state transport in the quantum spin Hall state}

\author
{Andreas Roth,$^{1}$ Christoph Br\"{u}ne,$^{1}$ Hartmut Buhmann,$^{1}$
Laurens W. Molenkamp,$^{1\ast}$\\
Joseph Maciejko,$^{2}$ Xiao-Liang Qi,$^{2}$
Shou-Cheng Zhang$^{2}$\\
\normalsize{$^{1}$Physikalisches Institut (EP3) and R\"{o}ntgen Center for Complex Material Systems,}\\
\normalsize{Universit\"{a}t W\"{u}rzburg, Am Hubland, 97074 W\"{u}rzburg, Germany}\\
\normalsize{$^{2}$Department of Physics, Stanford University, Stanford, CA 94305, USA}\\
\\
\normalsize{$^\ast$To whom correspondence should be addressed; E-mail:  molenkmp@physik.uni-wuerzburg.de.}
}

\date{}

\begin{document}


\baselineskip24pt


\maketitle


\begin{sciabstract}
We present direct experimental evidence for nonlocal transport
in HgTe quantum wells in the quantum spin Hall regime, in the absence of any
external magnetic field. The data conclusively show that the non-dissipative
quantum transport occurs through edge channels, while the contacts lead to
equilibration between the counter-propagating spin states at the edge. We
show that the experimental data agree quantitatively with the theory of the
quantum spin Hall effect.

\end{sciabstract}

The quantum spin Hall (QSH) state\cite{kane2005A,bernevig2006a} is
a topologically nontrivial state of matter which exists in the
absence of any external magnetic field. It has a bulk energy gap
but gapless helical edge states protected by time reversal
symmetry. In the QSH regime, opposite spin states forming a
Kramers doublet counter-propagate at the edge\cite{wu2006,xu2006}.
Recently, the QSH state has been theoretically predicted in HgTe quantum
wells\cite{bernevig2006d}. There is a topological
quantum phase transition at a critical thickness $d_c$ of the
quantum well, separating the trivial insulator state for $d<d_c$
from the QSH insulator state for $d>d_c$. Soon after the theoretical
prediction, evidence for the QSH state has been observed in transport measurements \cite{konig2007}.
In the QSH regime,
experiments measure a conductance close to $2e^2/h$, which is consistent with quantum transport
due to helical edge states. However, such a conductance quantization in small Hall bars does
not allow us to distinguish experimentally between ballistic and edge channel transport in
a convincing manner. Thus it is of the utmost importance for this field to be able to prove
experimentally in an unambiguous manner the existence of edge channels in HgTe quantum wells.

In conventional diffusive electronics, bulk transport satisfies Ohm's
law. The resistance is proportional to the length and inversely
proportional to the cross-sectional area, implying the existence
of a local resistivity or conductivity tensor. However, the
existence of edge states necessarily leads to nonlocal transport
which invalidates the concept of local resistivity. Such nonlocal
transport has been experimentally observed in the quantum Hall
(QH) regime in the presence of a large magnetic
field\cite{Beenakker1991}, and the nonlocal transport is well
described by a quantum transport theory based on the
Landauer-B\"{u}ttiker formalism\cite{Buttiker1988}. These measurements are now widely
acknowledged as constituting definitive experimental evidence for the existence of
edge states in the QH regime.

In this work, we report nonlocal transport measurements in HgTe quantum wells that
unequivocally demonstrate the existence of extended edge channels. We have fabricated
more complicated structures compared to a standard Hall bar that allow a detailed
investigation of the transport mechanism. The data present the first definitive
evidence for the actual occurrence of helical edge channels in our samples.
In addition, we present the theory of quantum transport in the QSH regime, and
uncover the remarkable effects of macroscopic time irreversibility on the helical edge states.

We present experimental results on four different devices, with
layouts as outlined below. The behavior in these structures  is
exemplary for the around 50 devices we studied. The devices
are fabricated from  HgTe/(Hg,Cd)Te quantum well (QW) structures
with well thicknesses of $d = 7.5$ nm (samples S1, S2 and S3) and
9.0 nm (sample S4). Note that all wells have a thickness $d > d_c
\simeq $ 6.3 nm, and thus exhibit the topologically non-trivial
inverted band structure.  At zero gate voltage, the samples are
n-type and have a carrier density of about $n_s = 3 \times
10^{11}$ cm$^{-2}$ and a mobility of $1.5 \times 10^{5}$
cm$^2$/(Vs), with small variations between the different wafers.
The actual devices are lithographically patterned using
electron-beam lithography and subsequent Ar ion-beam etching.
Devices S1 and S2 are micron-scale Hall bars with exact dimensions
as indicated in the insets of Fig. 1. S3 and S4 are dedicated
structures for identifying non-local transport, schematic
structure layouts are given in Fig.~2. All devices are fitted with
a 110-nm-thick Si$_3$N$_4$/SiO$_2$ multilayer gate insulator and a
5/50 nm Ti/Au gate electrode stack. By applying a voltage $V_g$ to
the top gate the electron carrier density of the QW can be
adjusted, going from an n-type behavior at positive gate voltages
through the bulk insulator state into a p-type regime at negative
gate voltages. For reasons of comparison, the experimental data in
Figs. 1,3, and 4 are plotted as a function of a normalized gate
voltage $V^{*}= V_ {g}-V_{thr}$ ($V_{thr}$ is defined as the
voltage for which the resistance is largest). Measurements are
performed at a lattice temperature of 10 mK using low-frequency
(13 Hz) lock-in techniques under voltage bias. The two terminal
and four terminal conductance results are shown in Fig. 1. The
four terminal resistance shows a maximum at about $h/2e^2$, in
agreement with the results of Ref. \cite{konig2007}. We also study
the two terminal resistance. The contact resistance should be
insensitive to the gate voltage, and can be measured from the
resistance deep in the metallic region. By subtracting the contact
resistance we find that the two terminal resistance has its
maximum of about $3h/2e^2$. As we shall see in the following
discussions, this value is exactly what is expected from the
theory of QSH edge transport obtained from the
Landauer-B\"{u}ttiker formula.

We now present the theory of quantum transport due to the helical
edge states in the QSH regime. Within the general
Landauer-B\"{u}ttiker formalism\cite{Buttiker1986}, the
current-voltage relationship is expressed as
\begin{equation}\label{Landauer}
I_i=\frac{e^2}{h} \sum_j (T_{ji} V_i - T_{ij} V_j),
\end{equation}
where $I_i$ is the current flowing out of the $i$-th electrode
into the sample region, $V_i$ is the voltage on the $i$-th
electrode, and $T_{ji}$ is the transmission probability from the
$i$-th to the $j$-th electrode. The total current is conserved in
the sense that $\sum_i I_i=0$. A voltage lead $j$ is defined by
the condition that it draws no net current, {\it i.e.} $I_j=0$.
The physical currents are left invariant if the voltages on all
electrodes are shifted by a constant amount $\mu$, implying that
$\sum_i T_{ij}=\sum_i T_{ji}$. In a time-reversal invariant
system, the transmission coefficients satisfy the condition
$T_{ij}=T_{ji}$.

For a general two-dimensional sample, the number of transmission
channels scales with the width of the sample, so that the
transmission matrix $T_{ij}$ is complicated and non-universal.
However, a tremendous simplification arises if the quantum
transport is entirely dominated by the edge states.
In the QH regime, chiral edge states are responsible for the
transport. For a standard Hall bar with $N$ current and voltage
leads attached (cf. the insets of Fig. 1 with $N=6$), the
transmission matrix elements for the $\nu=1$ QH state are given by
$T({\rm QH})_{i+1,i}=1$, for $i=1,\ldots,N$, and all other matrix
elements vanish identically. Here we periodically identify the
$i=N+1$ electrode with $i=1$. Chiral edge states are protected
from backscattering, therefore, the $i$-th electrode transmits
perfectly to the neighboring ($i+1$)th electrode on one side only.
In the example of current leads on the electrodes $1$ and $4$, and
voltage leads on the electrodes $2$, $3$, $5$ and $6$, one finds
that $I_1=-I_4\equiv I_{14}$, $V_2-V_3=0$ and
$V_1-V_4=\frac{h}{e^2} I_{14}$, giving a four-terminal resistance
of $R_{14,23}=0$ and a two-terminal resistance of
$R_{14,14}=\frac{h}{e^2}$.

In the case of helical edge states in the QSH regime, opposite
spin states form a Kramers pair, counter-propagating on the same
edge. The helical edge states are protected from backscattering
due to time reversal symmetry, and the transmission from one
electrode to the next is perfect. From this point of view, the
helical edge states can be viewed as two copies of chiral edge
states related by time reversal symmetry. Therefore, the
transmission matrix is given by $T({\rm QSH})=T({\rm
QH})+T^\dagger({\rm QH})$, implying that the only non-vanishing
matrix elements are given by
\begin{equation}\label{Transmission}
T({\rm QSH})_{i+1,i}=T({\rm QSH})_{i,i+1}=1,
\end{equation}
Considering again the example of current leads on the electrodes
$1$ and $4$, and voltage leads on the electrodes $2$, $3$, $5$ and
$6$, one finds that $I_1=-I_4\equiv I_{14}$,
$V_2-V_3=\frac{h}{2e^2} I_{14}$ and $V_1-V_4=\frac{3h}{e^2}
I_{14}$, giving a four-terminal resistance of
$R_{14,23}=\frac{h}{2e^2}$ and a two-terminal resistance of
$R_{14,14}=\frac{3h}{2e^2}$. The experimental data in Fig. 1
neatly confirm this picture. For both micro Hall-bar structures S1
and S2, that differ only in the dimensions of the area between the
voltage contacts 3 and 4 we observe exactly the expected
resistance values for $R_{14,23}=\frac{h}{2e^2}$ and
$R_{14,14}=\frac{3h}{2e^2}$ for gate voltages where the samples
are in the QSH regime.

Conceptually, one might sense a paradox between the
dissipationless nature of the QSH edge states and the finite
four-terminal longitudinal resistance
$R_{14,23}$, which vanishes for the QH state. 
We can generally assume that the microscopic Hamiltonian governing
the voltage leads is invariant under time reversal symmetry,
therefore, one would naturally ask how such leads could cause the
dissipation of the helical edge states, which are protected by
time reversal symmetry? In nature, the time reversal symmetry can
be broken in two ways, either at the level of the microscopic
Hamiltonian, or at the level of the macroscopic irreversibility in
systems whose microscopic Hamiltonian respects the time reversal
symmetry. When the helical edge states propagate without
dissipation inside the QSH insulator between the electrodes,
neither forms of time reversal symmetry breaking are present. As a
result, the two counter-propagating channels can be maintained at
two different quasi chemical potentials, leading to a net current
flow. However, once they enter the voltage leads, they interact
with a reservoir containing infinitely many low-energy degrees of
freedom, and the time reversal symmetry is effectively broken by
the macroscopic irreversibility. As a result, the two
counter-propagating channels equilibrate at the same chemical
potential, determined by the voltage of the lead. Dissipation
occurs with the equilibration process. The transport equation
(\ref{Landauer}) breaks the macroscopic time reversal symmetry,
even though the microscopic time reversal symmetry is ensured by
the relationship $T_{ij}=T_{ji}$. In contrast to the case of QH
state, the absence of dissipation of the QSH helical edge states
is protected by Kramers' theorem, which relies on the quantum
phase coherence of wavefunctions. Thus dissipation can occur once
the phase coherence is destroyed in the metallic leads. On the
contrary, the robustness of QH chiral edge states does not require
phase coherence. A more rigorous and microscopic analysis on the
different role played by a metallic lead in QH and QSH states is
provided in the supporting online text, the result of which agrees
with the simple transport equation (\ref{Landauer}) and
(\ref{Transmission}). These two equations correctly describe the
dissipationless quantum transport inside the QSH insulator, and
the dissipation inside the electrodes.
One can subject these two equations to more stringent experimental
tests than the two-and four-terminal experiments of Fig. 1 by
considering devices S3 and S4, as depicted in Fig. 2.

A further difference between helical and chiral edge channels is
evident from our experiments on the six-terminal device S3, as
shown in Fig. 3. When the longitudinal resistance  of device S3 is
measured by passing a current through contacts 1 and 4 and by
detecting the voltage between contacts 2 and 3 ($R_{14,23}$)
[Fig.~1a)], we find, similarly to the results of Fig. 1, the
celebrated resistance value of $h/2e^2$ when the bulk of the
device is gated into the insulating regime [Fig.~3 a)]. However,
the longitudinal resistance is significantly different in a
slightly modified
configuration, where the current is passed through contacts 1 and
3 and the voltage is measured between contacts 4 and 5
($R_{13,45}$) [Fig.~3 b)]. We now find $R_{13,45} \approx 8.6$
k$\Omega$ , which is markedly different from what one would expect
for either the QH transport, or the purely diffusive transport,
where this configuration would be equivalent to the
previous. Application of equations (\ref{Landauer}) and
(\ref{Transmission}) actually predicts indeed that the observed
behavior is what one expects for helical edge channels. One easily
finds that this resistance value can again be expressed as an
integer fraction of the inverse conductance quanta $e^2/h$:
$R_{13,45}= 1/3\, h/e^2$. This result shows that the current
through the device is influenced by the number of ohmic contacts
in the current path. As discussed earlier, these ohmic contacts
lead to the equilibration of the chemical potentials between the
two counter-propagating helical edge channels inside the contact.
There are also some devices for which the maximal resistance does
not match the theoretical value obtained from Eqs.
(\ref{Landauer}) and (\ref{Transmission}), but still remains an
integer fraction of the quantum $h/e^2$. This result can be
naturally understood as due to inhomogeneities in the gate action,
e.g. due to interface trap states, inducing some metallic droplets
close to the edge channels while the bulk of the sample is
insulating. A metallic droplet can cause dephasing of the
electronic wave function, leading to fluctuations in the device
resistance. For full dephasing, the droplet plays the role of an
additional Ohmic contact, just as for the chiral edge channels in
the QH regime\cite{Beenakker1991}. More details on the effects of
additional Ohmic contacts in the QSH state are given in the
supporting online text.

Another measurement that directly confirms the non-local character
of the helical edge channel transport in the QSH regime is in
Fig.~4, which shows data obtained from device S4, in the shape
of the letter ``H". In this 4-terminal device the current is
passed through contacts 1 and 4 and the voltage is measured
between contacts 2 and 3. In the metallic n-type regime (low gate
voltage) the voltage signal tends to zero. In the insulating
regime, however, the nonlocal resistance signal increases to
$\approx 6.5$ k$\Omega$, which again fits perfectly to the result
of Laudauer-B\"{u}ttiker considerations: $R_{14,23} =
h/4e^2\approx 6.45$ k$\Omega$. Classically, one would expect only
a minimal signal in this configuration (from Poisson's equation,
assuming diffusive transport, one estimates a signal of about 40
$\Omega$), and certainly not one that increases so strongly when
the bulk of the sample is depleted. This signal measured here is
fully non-local, and can be taken (as was done twenty years ago
for the QH regime) as definite evidence of the existence of edge
channel transport in the QSH regime. A similar non-local voltage
has been studied in a metallic spin Hall system with the same
H-bar geometry\cite{bruene2008}, in which case the nonlocal
voltage can be understood as a combination of the spin Hall effect
and the inverse spin Hall effect\cite{hankiewicz2004}. The
quantized nonlocal resistance $h/4e^2$ we find here is the quantum
counterpart of the metallic case. Assuming for example that the
chemical potential in contact 1 is higher than that in contact 4
(cf. the layout of S4 in Fig. 2 (b)), more electrons will be
injected into the upper edge state in the horizontal segment of
the H-bar than into the lower edge state. Since on opposite edges,
the right-propagating edge states have opposite spin, this implies
that a spin-polarized current is generated by an applied bias
$V_1-V_4$, comparable to a spin Hall effect. When this
spin-polarized current is injected into the right leg of the
device, the inverse effect occurs. Electrons in the upper edge
flow to contact 2 while those in the lower edge will flow to
contact 3, establishing a voltage difference between those two
contacts due to the charge imbalance between the edges. The right
leg of the device thus acts as a detector for the injected
spin-polarized current, which corresponds to the inverse spin Hall
effect.

In conclusion, we have shown multi-terminal and non-local
transport experiments on HgTe microstructures in the QSH regime
that unequivocally demonstrate that charge transport occurs
through extended helical edge channels. We have extended the
Landauer-B\"{u}ttiker model for multi-terminal transport in the QH
regime to the case of helical QSH edge channels, and have shown
that this model convincingly explains our observations. These
results constitute decisive evidence that the
conductance quantization observed in Ref.\cite{konig2007} stems
from QSH edge channel transport, which may
be used for non-dissipative transfer of information.


\bibliography{NonlocalPreprint}

\bibliographystyle{Science}


\begin{scilastnote}
\item We thank T. Beringer, N. Eikenberg, M. K\"{o}nig and
  S. Wiedmann for assistance in some of the experiments. We
  gratefully acknowledge financial support by the Deutsche
   Forschungsgemeinschaft (SFB 410), the German-Israeli Foundation
   (I-881-138.7/2005, the NSF (grant DMR-0342832), the U.S.
   Department of Energy, Office of Basic Energy Sciences, under
    contract DE-AC03-76SF00515, the Focus Center Research Program
    (FCRP) Center on Functional Engineered Nanoarchitectonics
    (FENA), the National Science and Engineering
    Research Council (NSERC) of Canada, and the Stanford Graduate Fellowship
    Program (SGF). Computational work was made possible by the facilities of the Shared
Hierarchical Academic Research Computing Network
(SHARCNET:www.sharcnet.ca).
\end{scilastnote}



\clearpage

\begin{figure}[h]
\begin{center}
\includegraphics[width=6in]{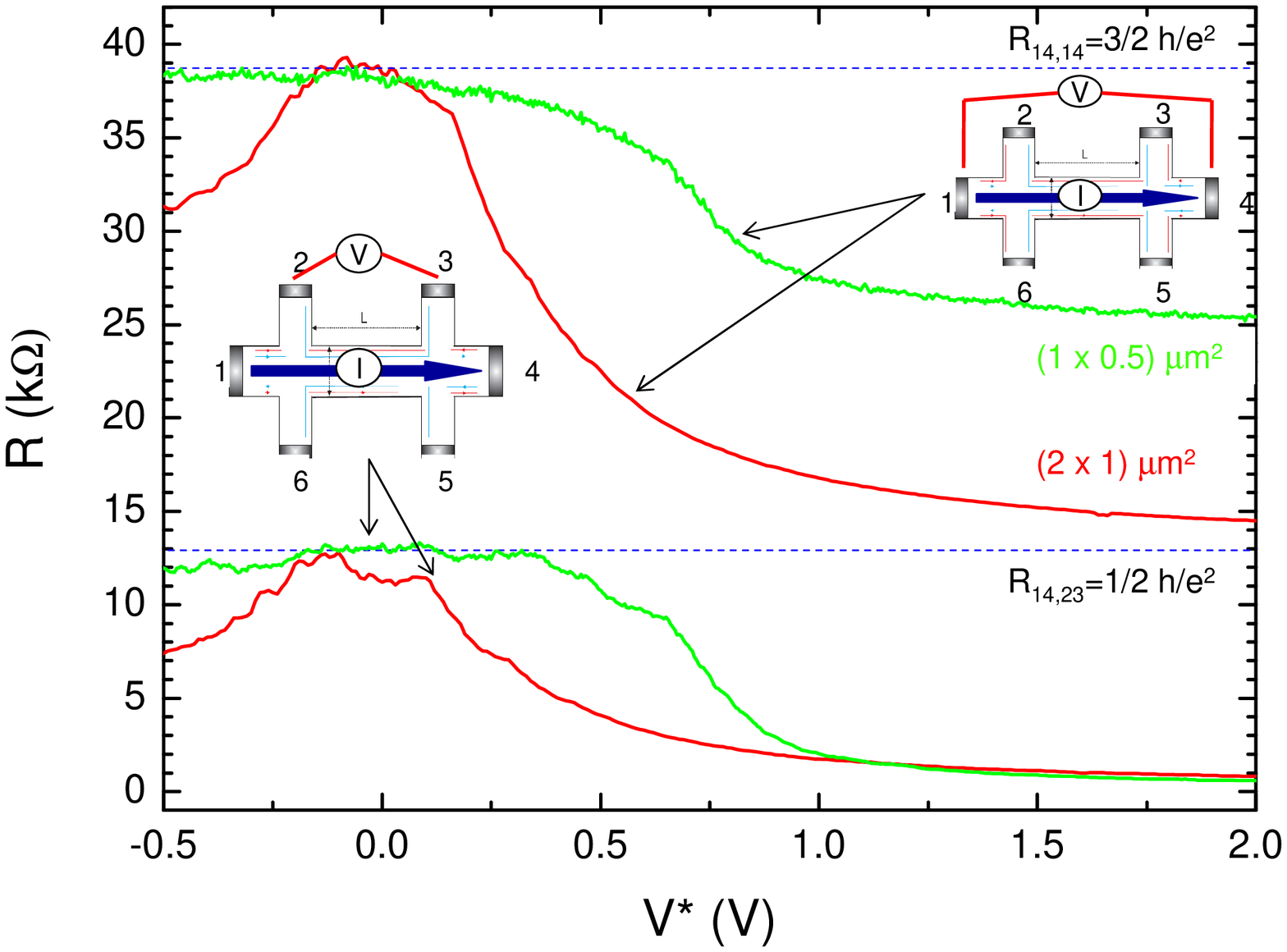}
\end{center}
\end{figure}

\noindent {\bf Fig. 1.} Two-terminal ($R_{14,14}$) and four-
terminal ($R_{14,23}$) resistance versus (normalized) gate voltage
for the Hall bar devices S1 and S2 with dimensions as shown in the
insets. The dotted blue lines indicate the resistance values
expected from the Landauer-B\"{u}ttiker approach.

\clearpage

\begin{figure}[h]
\begin{center}
\includegraphics[width=6in]{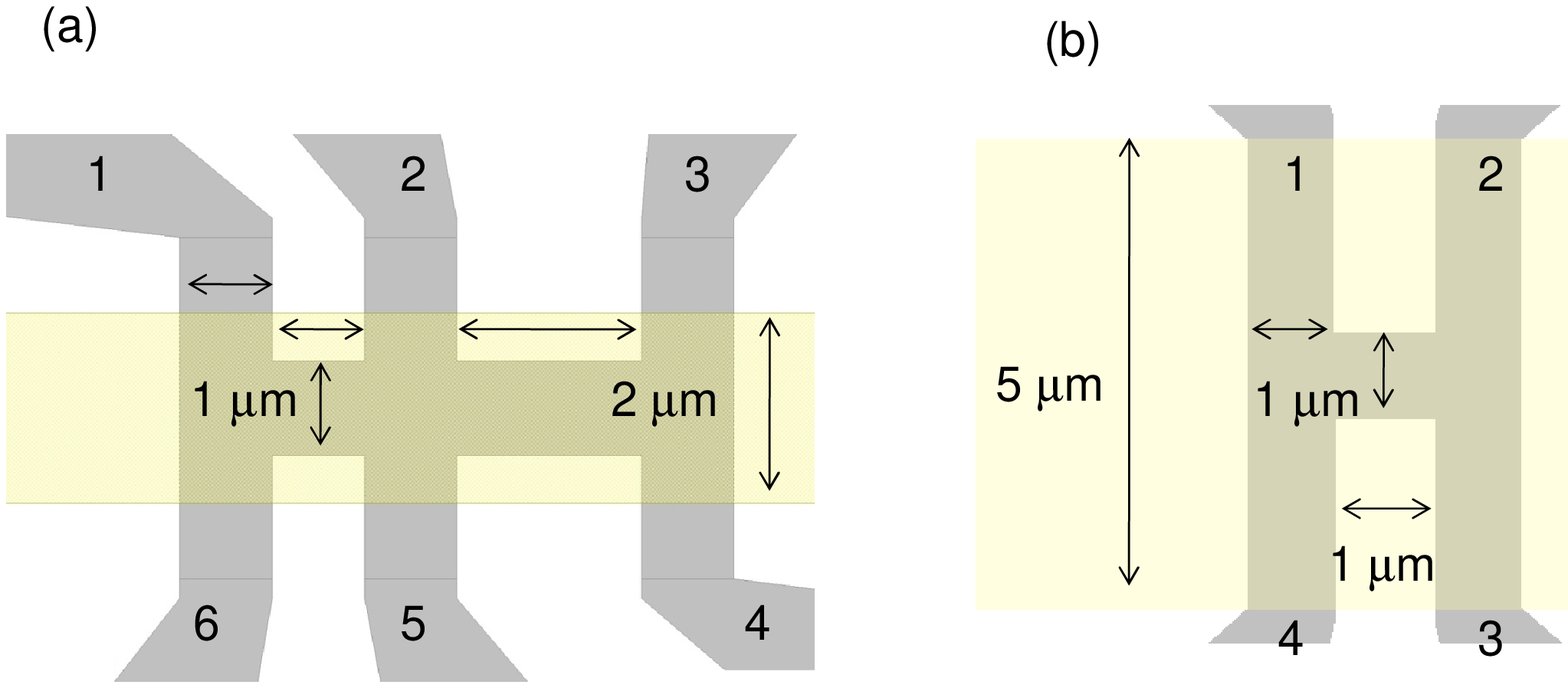}
\end{center}
\end{figure}

\noindent {\bf Fig. 2.} Schematic layout of  devices S3 (a) and S4
(b). The grey areas are the mesa's, the yellow areas the gates,
with dimensions as indicated in the figure. The numbers indicate
the coding of the leads.

\clearpage

\begin{figure}[h]
\begin{center}
\includegraphics[width=6in]{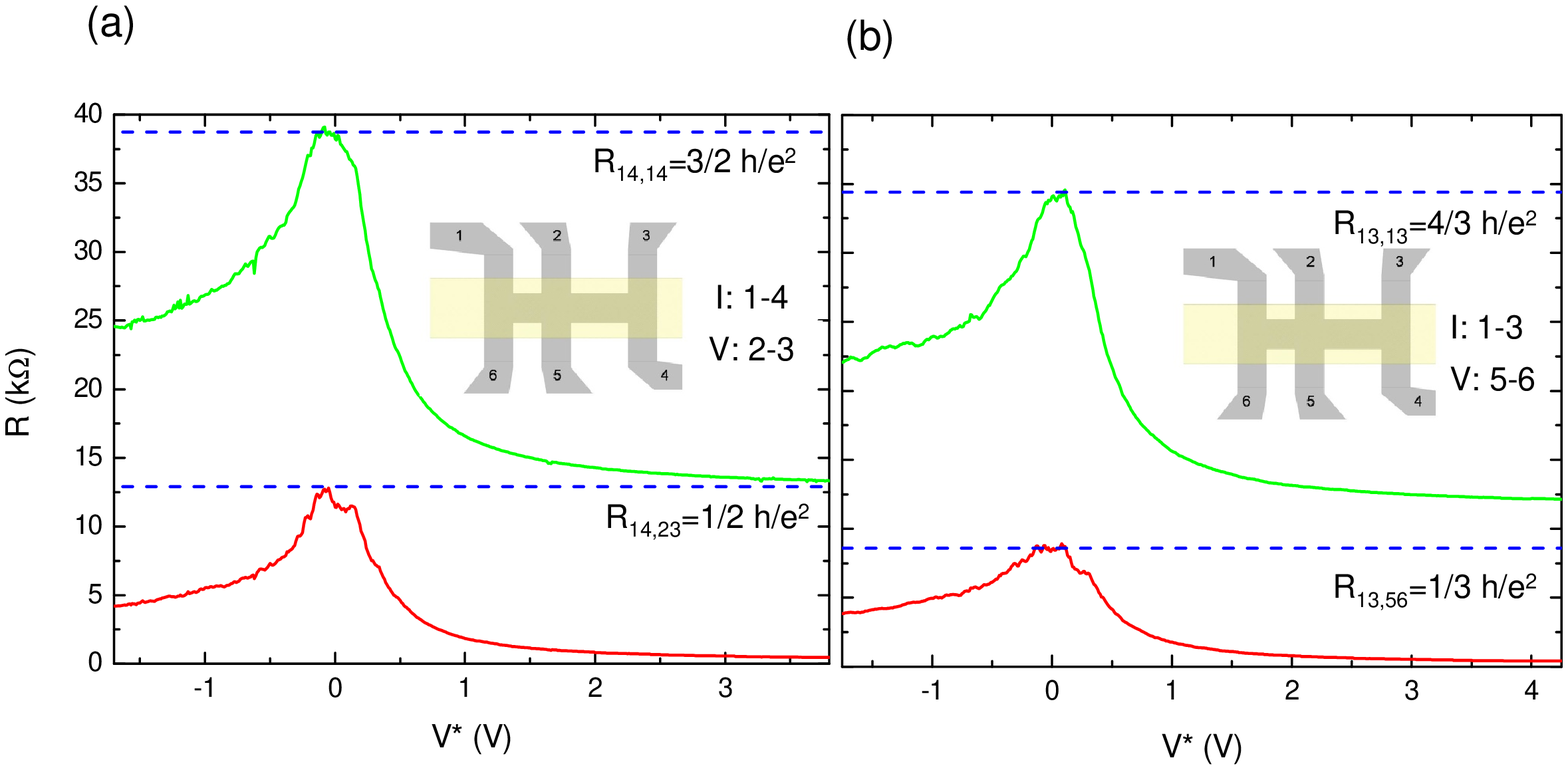}
\end{center}
\end{figure}

\noindent {\bf Fig. 3.} Four- and two-terminal resistance measured
on device S3: (a) $R_{14,23}$ (red line) and $R_{14,14}$ (green
line) and (b) $R_{13,56}$ (red line) and $R_{13,13}$ (green line).
The dotted blue lines indicate the expected resistance value from
a Landauer-B\"uttiker calculation.

\clearpage

\begin{figure}[h]
\begin{center}
\includegraphics[width=6in]{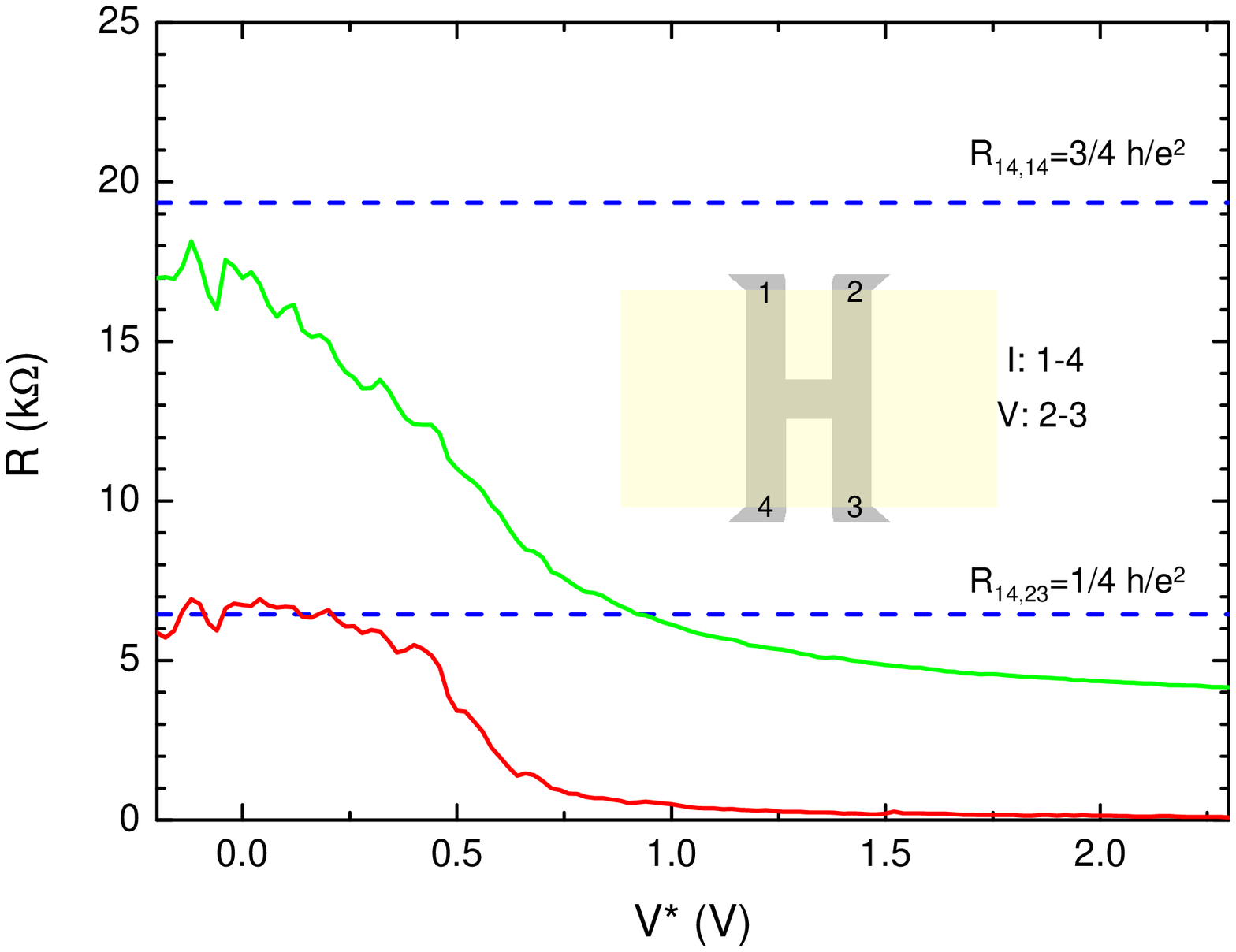}
\end{center}
\end{figure}

\noindent {\bf Fig. 4.} Nonlocal four-terminal resistance and
two-terminal resistance measured on the H-bar device S4:
$R_{14,23}$ (red line) and $R_{14,14}$ (green line). Again, the
dotted blue line represents the theoretically expected resistance
value.

\end{document}